\documentclass[twocolumn]{jpsj2}
\title{Diversity in Free Energy Landscape of Proteins with the Same Native Topology}
\author{Hiroo \textsc{Kenzaki}$^{1,2,3}$\thanks{E-mail address: kenzaki@tbp.cse.nagoya-u.ac.jp}
and Macoto \textsc{Kikuchi}$^{2,1}$\thanks{E-mail address: kikuchi@cmc.osaka-u.ac.jp}}
\inst{
$^{1}$Department of Physics, Osaka University, Toyonaka 560-0043\\
$^{2}$Cybermedia Center, Osaka University, Toyonaka 560-0043\\
$^{3}$JST-CREST, Nagoya 464-8601}

\abst{In order to elucidate the role of the native state topology and the stability of subdomains in protein folding, 
we investigate free energy landscape of human lysozyme,
which is composed of two subdomains, by Monte Carlo simulations.
A realistic lattice model with G\={o}-like interaction is used.
We take the relative interaction strength (stability, in other word) 
of two subdomains as a
variable parameter and study the folding process.
A variety of folding process is observed and we obtained a
phase diagram of folding in terms of temperature and the relative stability.
Experimentally-observed diversity in folding process of c-type lysozimes
is thus understood as a consequence of the difference in the relative stability
of subdomains.
}
\kword{protein folding, free energy landscape, native topology, lysozyme, multicanonical ensemble, G\={o} model}

\begin{document}
\maketitle
%%%%%%%%%%%%%%%%%%%%%%%%%%%%%%%%%%%%%%%%%%%%%%%%%%%%%%%%%%%%%%%%%%%%%%%%%%
Proteins take particular conformations called the native states under 
the physiological condition.
How proteins fold into their native states through vast
conformational spaces has been a fundamental problem in biophysics.
Recent development of the energy landscape theory has made it 
more and more evident that the folding processes are determined largely
by the chain conformations of the native states  (often called as \textit{native topology}).
According to the theory, conformation spaces of proteins become narrower towards the native states and the resulted energy landscapes are said symbolically as \textit{funnel-like shaped}.\cite{onuchic97,pande00} 
In other word, 
proteins are designed so that the energetical frustration is minimized.

Rise and development of the energy landscape theory has induced the recent revival of G\={o} model,\cite{taketomi75}
which includes interactions only between amino acids that contact in the native state (such an amino acid pair is called as a \textit{native pair}) and thus the native state is automatically the ground state.
Actually, G\={o} model is considered as a minimalistic realization of the funnel-like energy landscape.
Some variants of G\={o} model
have been shown to reproduce various properties of protein folding.
\cite{clementi00,koga01}
Remarkably, 
validity of G\={o}-like models is not limited only to 
the proteins that exhibit simple folding processes;
some proteins having intermediate states in the folding processes
are also described well by G\={o}-like models.
Recent success of G\={o}-like models supports the abovementioned postulate that  the native topology is the most important factor that determines the folding process, since they are based on the information only of the native state.
In fact, proteins having a similar native structure follow a similar folding path in many cases.
Some proteins with the same native topology, however, are known to fold via different pathways.\cite{zarrine05}
Thus, it is true that
the folding process is governed largely by the native topology,
but there are some cases that other factors dominate.

Following the statistical concept of the energy landscape theory,
the folding route of a protein is determined by the free energy
of the transition state and, if exists, the intermediate state.
However, analysis of folding process along this line has been limited mainly to single-domain proteins.
Validity of the G\={o}-like models as well as the funnel picture of the energy landscape for larger multi-domain proteins is still an open question.
Considering the fact that most of the real proteins consist of two or more subdomains, understanding the folding processes of multi-domain proteins is an important subject.
In this letter, we investigate folding process of proteins of two subdomains 
within the framework of G\={o}-like model.
In particular, we focus on influence of inhomogeneity in energetic stability of subdomains.

\begin{figure}[tb]
\begin{center}
\includegraphics{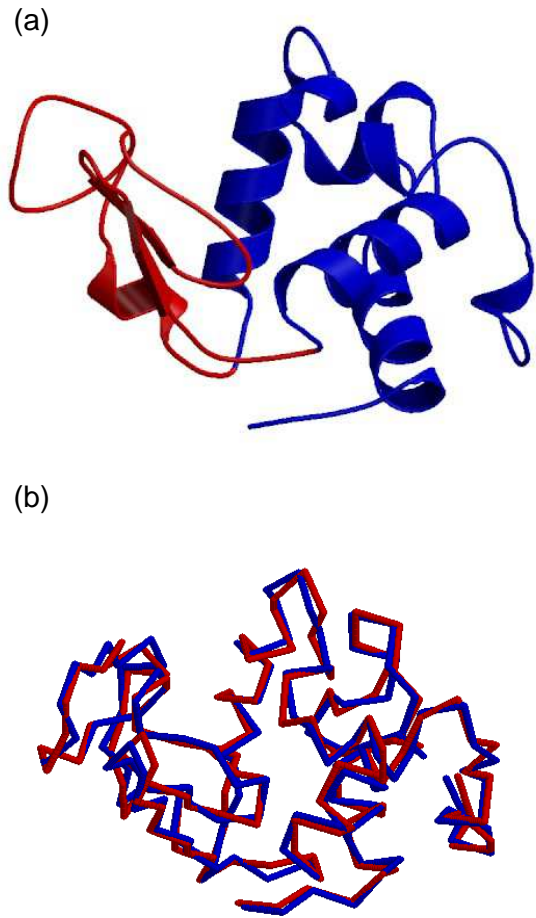}
\caption{(a)X-ray crystal structure of human lysozyme.
Blue indicates $\alpha$ domain (residues 1-38 and 88-130) 
and red indicates $\beta$ domain (residues 39-87).
This image was prepared using 
MOLSCRIPT\cite{kraulis91} and Raster3D\cite{merritt97}.
(b) Superposition of the C$^{\alpha}$ trace of X-ray crystal structure of human lysozyme (red) and its lattice realization (blue)}
\label{fig1}
\end{center}
\end{figure}

We deal with chicken-type (c-type) lysozyme, a class of two-domain proteins, in which a number of proteins with a similar native structure are included such as several variants of lysozymes and $\alpha$-lactalbumin.
Folding processes of c-type lysozymes have been extensively investigated by experiments, because they are relatively small among multi-domain proteins.
In fact, it is currently one of the {\it standard} materials for study of the folding processes of multi-domain proteins.
The c-type lysozymes have two subdomains:
$\alpha$ subdomain, which is composed of two helical regions both of N-terminal and C-terminal,
and $\beta$ subdomain, which is the middle beta region[Fig. \ref{fig1}(a)].
Many proteins in this class exhibit three-state folding kinetics;
in other words, intermediate states are observed in the folding processes.
Nature of the intermediate states are different from protein to protein.
Hen lysozyme and human lysozyme 
 have intermediate states in the folding processes (kinetic intermediates), 
while no corresponding equilibrium state (equilibrium intermediate) 
has been observed under sweeping of external parameters such as temperature.\cite{radford92,hooke94}
Thus, the intermediates states exist only as thermodynamically metastable states in these cases.
On the other hand, canine milk lysozyme and $\alpha$-lactalbumin have equilibrium intermediates\cite{koshiba00,schulman97}.
Considering the native topology, we intuitively expect that
$\beta$ subdomain will fold first, and then $\alpha$ subdomain, because formation of $\beta$ subdomain requires contacts of amino acids that are distantly located along the chain.
A G\={o}-like model with homogeneous interactions predicts the same folding route as will be shown later.
But the experimental evidences indicate opposite;
$\alpha$ domain already folded partially both in the kinetic intermediates and the equilibrium ones, while $\beta$ domain is still disordered there.

Interactions in G\={o}-like models are often taken to be homogeneous, that is, 
the same interaction parameter is applied to all the native pairs.
Although such setting is the simplest choice, 
we are allowed also to use different interaction parameters for different pairs within the framework of G\={o}-like models, if only the native pairs are taken into account.
In fact, G\={o}-like models using Miyazawa--Jernigan type heterogeneous interactions\cite{miyazawa96} have also been used,\cite{karanicolas03,ejtehadi04,das05} considering that interactions between amino acid residues are heterogeneous in reality.
In this letter, however, we propose a totally different treatment for the interaction parameters.
Instead of fixing the interaction parameters to particular values, we regard them as variable parameters, and thereby we discuss diversity of possible folding routes within proteins of the same topology in a unified manner.

%%%%%%%%%%%%%%%%%%%%%%%%%%%%%%%%%%%%%%%%%%%%%%%%%%%%%%%%%%%%%%%%%%%%%%%%%%%%

It is widely recognized that a folding simulation of a protein is a difficult task even with today's high performance computers.
Calculation of the global free energy landscape should be even more difficult,
because we need to sample conformations of wide energy range with accurate weight.
Lattice models are favorable for this purpose from the view point of computational cost.
Some realistic lattice models
that can represent flexible structures of real proteins in a satisfactory level of approximation have been proposed.\cite{kolinski04}
In the present work, we use the 210-211 hybrid lattice model
in which an amino acid residue is represented by its C$^{\alpha}$ atom located on a simple cubic lattice;
All other atoms such as N and H are not explicitly considered and the side chain as well.
Consecutive C$^{\alpha}$ atoms are connected by vectors
of the type (2,1,0) or (2,1,1) and all the possible permutations.
In order to take into account the excluded volume effect of amino acid residues, we assume that each amino acid occupies seven lattice sites,
that is, a center site where C$^{\alpha}$ atom is located and all of its nearest-neighbor sites.
Construction of G\={o}-like model requires knowledge about the native structure.
For that purpose, we take the X-ray crystal structure of human lysozyme (Protein Data Bank Code: 1jsf), which consists of 130 amino acid residues, as a reference structure;
then the native structure of the lattice protein model is obtained by
fitting to its C$^{\alpha}$ trace.
Figure \ref{fig1}(b) shows the native structure of the lattice model thus obtained and that of human lysozyme.
Their root mean square deviation (rmsd) is $0.85$ \AA, 
which we regard is satisfactorily accurate for the present purpose.

We introduce G\={o}-like interactions which act only in the native pairs.
A harmonic-type local interaction is also introduced, which expresses the excess energy due to the bond stretching.
Then the
\textit{Hamiltonian} for the \textit{homogeneous} case is defined as follows:

\begin{align}
V_h & = \frac{K_b}{2} \sum_{i=1} (r_{i,i+2} - r_{i,i+2}^{nat})^2- \sum_{j-i>2}
\varepsilon C_{i,j} \Delta (r_{i,j}, r_{i,j}^{nat}),\\
\Delta(x,y) & = \left\{
\begin{array}{lc}
1 & |x^2 - y^2| \le W\\
0, & \textrm{otherwise}
\end{array} \right.
\end{align}
\noindent
where $i$ and $j$ indicate the residue number counted along the chain, $r_{i,j}$ is the distance between C$^{\alpha}$ atoms of $i$-th and $j$-th residues, 
$r_{i,j}^{nat}$ is their native distance,
$W$ is the width of the G\={o} potential,
$K_{b}$ is strength of the local interaction,
$\varepsilon$ is strength of the G\={o} potential
and $C_{i,j}$ is assigned a value $1$ or $0$ 
depending on whether the pair is the native pair or not.
We use $W = 2$, $K_b = 1$ and $\varepsilon = 1$ throughout this work.
A pair of residues is considered as a native pair
if the minimum distance between their heavy atoms
is less than $4.5$ \AA{} in the X-ray crystal structure.

Then we introduce inhomogeneity in interactions.
For the purpose of investigating the effect of subdomain stability,
we divide the interaction $V_h$ into two parts, the interactions within the $\alpha$ domain and the rest, and write the new potential as $V = R_{\alpha} V_{\alpha} + R_{\beta} V_{\beta}$, where $R_{\alpha}$ and $R_{\beta}$ are variable parameters.
We should note that inter-subdomain interaction is included in $V_{\beta}$.
For efficient sampling of conformations,
we employ a variant of two-energy multicanonical ensemble Monte Carlo method,\cite{berg92,lee93,iba98,chikenji99}
in which two dimensional histogram in ($V_{\alpha}$, $V_{\beta}$)--space is made to be flat.
This method enables us to study equilibrium properties for any values of $R_{\alpha}$ and $R_{\beta}$ through the reweighting procedure.

%%%%%%%%%%%%%%%%%%%%%%%%%%%%%%%%%%%%%%%%%%%%%%%%%%%%%%%%%%%%%%%%%%%%%%%%%%%%

\begin{figure}[tb]
\begin{center}
\includegraphics{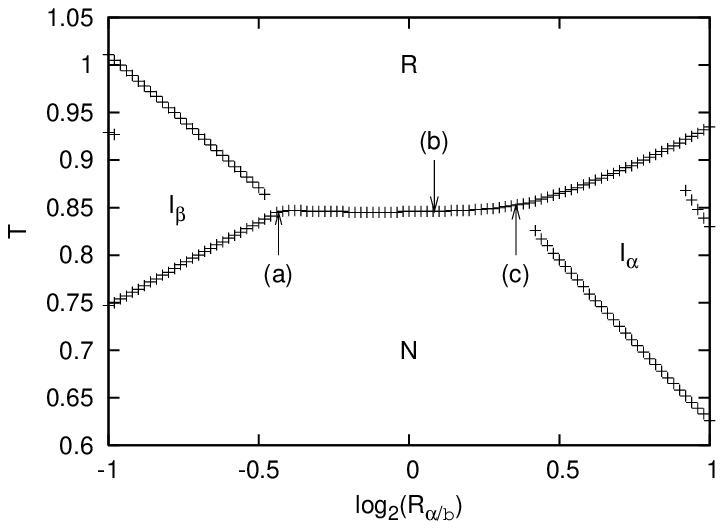}
\caption{Phase diagram of folding of lysozyme.
Transition points, estimated from the peaks of heat capacity,
are shown for various $R_{\alpha/\beta}$.
N indicates the native state and R is for the random coil.
I$_{\alpha}$ is an intermediate state with $\alpha$ domain formed,
and I$_{\beta}$ is an intermediate state with $\beta$ domain formed.
Values of $R_{\alpha/\beta}$ and $T$ indicated by the arrows are:
(a) $R_{\alpha/\beta} = 0.74$ and $T=0.845$,
(b) $R_{\alpha/\beta} = 1.06$ and $T=0.846$,
(c) $R_{\alpha/\beta} = 1.28$ and $T=0.853$.}
\label{fig2}
\end{center}
\end{figure}

We investigate how does
the weight parameter $R_{\alpha/\beta}\equiv R_{\alpha}/R_{\beta}$
influences the folding behavior.
Figure \ref{fig2} shows the phase diagram in terms of temperature $T$ and $R_{\alpha/\beta}$.
The total energy of the native state is fixed same.
We define the transition point
as temperature that the heat capacity takes a (local) maximal value.
We found that number of transitions varies between one and three 
in the range of $0.5 \le R_{\alpha/\beta} \le 2$.
Lysozyme folds into the native state N below the lowest transition temperature and unfolds to be a random coil R above the highest transition temperature
irrespective of the value of $R_{\alpha/\beta}$
(these two transitions coincide in the single transition case)
For $R_{\alpha/\beta} \le 0.74$, an equilibrium intermediate state is observed in a finite temperature range, which we call as  I$_\beta$.
Another equilibrium intermediate state, which we call as I$_\alpha$, is observed for $R_{\alpha/\beta} \ge 1.28$.
An additional weak transition is found within the intermediate state region
when $R_{\alpha/\beta}$ close to $0.5$ or $2$;
peak of the heat capacity is relatively low at these additional transitions
and thus only a slight structural change is considered to take place;
we will not pursue this transition further.

\begin{figure}[tb]
\begin{center}
\includegraphics{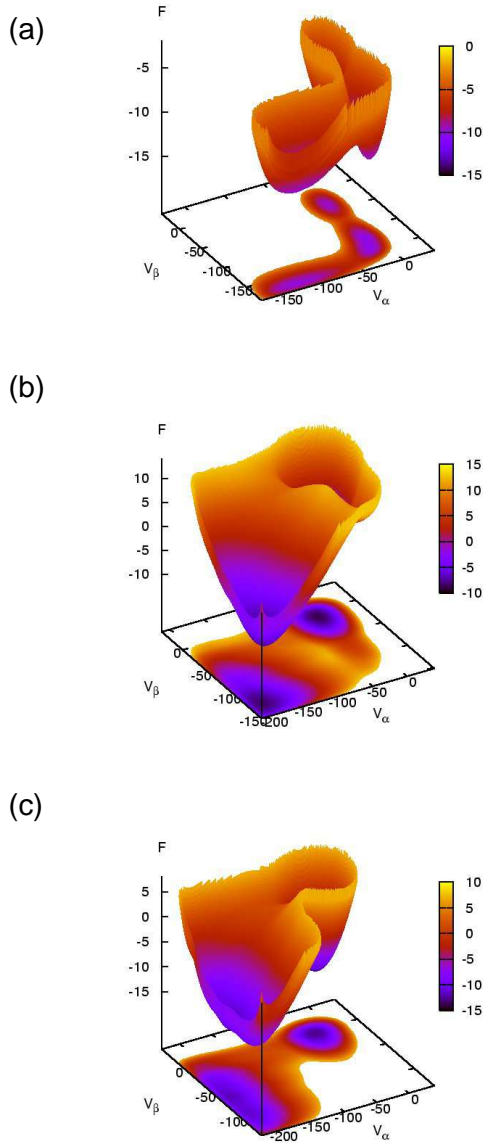}
\caption{Free energy landscape of lysozyme
as a function of $V_\alpha$ and $V_\beta$.
(a) $R_{\alpha/\beta} = 0.74$ and $T=0.845$,
(b) $R_{\alpha/\beta} = 1.06$ and $T=0.846$,
(c) $R_{\alpha/\beta} = 1.28$ and $T=0.853$.}
\label{fig3}
\end{center}
\end{figure}

Figure \ref{fig3}(a)-(c) shows the free energy landscapes
at three selected points (a)-(c) indicated by the allows in Fig. \ref{fig2}, respectively.
Three minima corresponding to R, I$_{\beta}$ and N are observed at (a).
The free energy landscape clearly shows that 
$\beta$ subdomain forms in the intermediate states I$_\beta$.
Three minima are also seen at (b), which in this case correspond to R, I$_{\alpha}$ and N.
$\alpha$ subdomain forms in the intermediate state I$_\alpha$.
Thus different folding pathways are followed in these two cases.
Two pathways between R and N have the same weight at (c), which are divided by a free energy barrier.
In contrast to the above two cases, neither of two pathways contains a metastable state in this case, so that the folding at this point should be of two-state type without a kinetic intermediate state.
Considering also the free energy landscapes for other values of $R_{\alpha/\beta}$ (figures not shown), the folding behavior at the lower transition point (folding temperature) is summarized as follows:
(i)For $R_{\alpha/\beta} \le 0.74 (a)$,
I$_\beta$ is the equilibrium intermediate state.
Thus it also is a kinetic intermediate at the folding temperature.
(ii)I$_\beta$ becomes a metastable state as $R_{\alpha/\beta}$ increases from $0.74$, and thus I$_\beta$ acts as a kinetic intermediate.
(iii)the metastable state disappears at some $R_{\alpha/\beta}<1.06$ (b).
The folding becomes two-state type and $beta$ subdomain forms first in the principal pathway.
(iv)Two-state folding with two equally possible pathway at (c).
(v)Two-state folding with $alpha$ subdomain form first in the principal pathway for $R_{\alpha/\beta}>1.06$.
(vi)the metastable state corresponding to I$_\alpha$ appears at some $R_{\alpha/\beta}>1.06$. I$_\beta$ acts as a kinetic intermediate.
(vii)For $R_{\alpha/\beta} > 1.28 (c)$,
I$_\alpha$ is the equilibrium intermediate state and thus also is a kinetic intermediate.

Change in a single parameter $R_{\alpha/\beta}$ thus results in a wide spectrum of folding behavior.
The cases (i) and (vii) may be expected naturally from the energy balance,
but existence of the finite region of two-state folding with two folding pathways is rather a nontrivial result.
%Change of the main folding route in the two-state region si 
The homogeneous G\={o}-like model $R_{\alpha/\beta} = 1$ is included in the case (iii), while the experimental evidence suggests that
the human lysozyme actually corresponds to the case (vi)\cite{hooke94}.
This discrepancy indicates that the $\alpha$ subdomain has high stability
compared to the $\beta$ subdomain in human lysozyme.
Differences found in folding behavior of other c-type lysozymes,
such as existence or nonexistence of an equilibrium intermediate state,
can also be understood as a result of difference in $R_{\alpha/\beta}$.
This postulate is further supported by an experimental evidence that 
the human lysozyme becomes to have an equilibrium intermediate state
when a part of the protein is replaced by the corresponding part of $\alpha$-lactalbumin;\cite{pardon95}
in the present context, the human lysozyme seems to shift from the case (vi) to the case (vii) due to the change of interaction parameters.

Lysozyme is not the only example of proteins that a subdomain 
or, for smaller proteins, a folding unit 
composed both of N-terminal and C-terminal regions rather than the central region is formed first in the equilibrium intermediate state .\cite{li99}
We expect that the results obtained above can be generalized for understanding the folding mechanism of these proteins, because existence of the cases (i) and (vii) is considered to be a natural consequence of two domains.
In this view point, the folding processes of the above proteins correspond to the case (vii).
Of course, details such as the number of pathways in the two-state transition region will differ largely from protein to protein.
Many experimental evidences have also been reported that
the folding processes can be modified by artificial mutation methods
leaving the native topology unchanged.\cite{zarrine05}
%Mutation of only one residue sometimes change the folding process
%between two-state (without a folding intermediate) and three-state (with a folding intermediate) kinetics.
Although these experiments are mainly for single-domain proteins, 
the mechanism of these changes may be understood as a consequences of 
the relative interaction strength of folding units, which are smaller units than a subdomain.

%To summarize, 
%we investigated the free energy landscape that describes the folding of lysozyme using a G\={o}-like model on a simple cubic lattice.

To summarize,
we observed a variety of folding processes of lysozyme
by varying the relative strength of the interactions in two subdomains
in spite that the native conformation is kept unchanged.
This result suggests that the experimentally observed diversity in folding behavior of c-type lysozymes can be understood as a consequence of difference in the subdomain stability.
It should be stressed that the model is still within the framework of G\={o}-like model.
Thus, a variety of folding route for proteins of the same native topology can be understood by the funnel picture.
We expect that folding sequences of multi-domain proteins are in general influenced by subdomain stability as well as the native topology.
In order to describe such a variation of folding processes in a unified manner, 
G\={o}-like model with variable interaction strength will be usefully.

%%%%%%%%%%%%%%%%%%%%%%%%%%%%%%%%%%%%%%%%%%%%%%%%%%%%%%%%%%%%%%%%%%%%%%%%%%%%%

\section*{Acknowledgment}
We would like to thank M. Sasai, S. Takada and G. Chikenji for critical reading of the manuscript and valuable comments.
The present work is partially supported by IT-program of Ministry of Education,
Culture, Sports, Science and Technology, The 21st Century COE program named
"Towards a new basic science: depth and synthesis", Grant-in-Aid for Scientific
Research (C) (17540383) from Japan Society for the Promotion of Science
and JST-CREST.


\begin{thebibliography}{99}
\bibitem{onuchic97} J.N. Onuchic, Z. Luthey-Schulten and P.G. Wolynes: Anuu. Rev. Phys. Chem. \textbf{48} (1997) 545.
\bibitem{pande00} V.S. Pande, A.Y. Grosberg and T. Tanaka: Rev. Mod. Phys. \textbf{72} (2000) 259.
\bibitem{taketomi75} H. Taketomi, Y. Ueda and N. G\={o}: Int. J. Peptide Protein Res. \textbf{7} (1975) 445.

\bibitem{clementi00} C. Clementi, H. Nymeyer and J.N. Onuchic: J. Mol. Biol. \textbf{298} (2000) 937.
\bibitem{koga01} N. Koga and S. Takada: (2001) J. Mol. Biol. \textbf{313} (2001) 171.
\bibitem{zarrine05} A. Zarrine-Afsar, S.M. Larson and A.R. Davidson: (2005) Curr. Opin. Struct. Biol. \textbf{15} (2005) 42.

\bibitem{radford92} S.E. Radford, C.M. Dobson and P.A. Evans: Nature \textbf{358} (1992) 302.
\bibitem{hooke94} S.D. Hooke, S.E. Radford and C.M. Dobson: Biochemistry \textbf{33} (1994) 5867.
\bibitem{koshiba00} M. Kikuchi, K. Kawano and K. Nitta: Protein Sci. \textbf{7} (1998) 2150.
\bibitem{schulman97} B.A. Schulman, P.S. Kim, C.M. Dobson and C. Redfield: Nature Struct. Biol. \textbf{4} (1997) 630.

%\bibitem{onuchic97} %(arpc97.pdf)
%\bibitem{gunasekaran01} K. Gunasekaran, S.J. Eyles, A.T. Hagler and L.M. Gierasch: Curr. Opin. Struct. Biol. \textbf{11} (2001) 83.

%\bibitem{dinner00} A.R. Dinner, A. \v{S}ali, L.J. Smith, C.M. Dobson and M. Karplus: Trends Biol. Sci. \textbf{25} (2000) 331.
\bibitem{miyazawa96} S. Miyazawa and R.L. Jernigan: J. Mol. Biol. \textbf{256} (1996) 623.
\bibitem{karanicolas03} J. Karanicolas and C.L. Brooks III: J. Mol. Biol. \textbf{334} (2003) 309.
\bibitem{ejtehadi04} M.R. Ejtehadi, S.P. Avall and S.S. Plotkin: Proc. Natl. Acad. Sci. USA \textbf{101} (2004) 15088.
\bibitem{das05} P. Das, S. Matysiak and C. Clementi: Proc. Natl. Acad. Sci. USA \textbf{102} (2005) 10141.
\bibitem{kolinski04} A. Kolinski and J. Skolnick: Polymer \textbf{45} (2004) 511.
\bibitem{kraulis91} P.J. Kraulis: J. Appl. Crystallogr. \textbf{24} (1991) 946.
\bibitem{merritt97} E.A. Merritt and D.J. Bacon, Methods Enzymol. \textbf{277} (1997) 505.
\bibitem{berg92} B.A. Berg and T. Neuhaus: Phys. Rev. Lett. \textbf{68} (1992) 9.
\bibitem{lee93} J. Lee: Phys. Rev. Lett. \textbf{71} (1993) 211.
\bibitem{iba98} Y. Iba, G. Chikenji and M. Kikuchi: J. Phys. Soc. Jpn. \textbf{67} (1998) 3327.
\bibitem{chikenji99} G. Chikenji, M. Kikuchi and Y. Iba: Phys. Rev. Lett. \textbf{83} (1999) 1886.
\bibitem{pardon95} E. Pardon, P. Haezebrouck, A. De Baetselier, Shaun D. Hooke, K.T. Fancourt, J. Desmet, C.M. Dobson, H. Van Dael and M. Joniau: J. Biol. Chem. \textbf{270} (1995) 10514.
\bibitem{li99} R. Li and C. Woodward: Protein Sci. \textbf{8} (1999) 1571.
\end{thebibliography}
\end{document}